\documentclass[12pt]{iopart}
\usepackage{times}
\usepackage{iopams}
\usepackage{amssymb}
\usepackage{float}
\usepackage[dvips]{graphicx}
\begin{document}
\title{Squeezing of a coupled state of two spinors }
\author{A R Usha Devi\dag, K S Mallesh\ddag, Mahmoud A A Sbaih\ddag, K B Nalini\S\  and\\
G Ramachandran*}
\address{\dag\ Department of Physics, Bangalore University,
Bangalore 560 056,India}
\address{\ddag\ Department of Studies in Physics, University of Mysore,
Mysore 570 006, India}
\address{\S Vidya  Vikas Institute of Engineering and Technology,
Mysore 570 010, India }
\address{* Indian Institute of Astrophysics, Koramangala, Bangalore 560 034,
India } 
 
\begin{abstract}
The notion of spin squeezing involves reduction in the uncertainty of a
component of the spin vector ${\vec S}$ below a certain limit.  This aspect has been studied
earlier \cite{meh1,meh2} for pure and mixed states of definite spin.
In this paper, this study has been extended to coupled spin states which
do not possess sharp spin value.  A general squeezing criterion has been obtained by requiring that 
a direct product state for two spinors is not squeezed. The squeezing aspect of entangled 
states is studied in relation to their spin- spin correlations. 
\end{abstract}

\submitted
\maketitle

\section{Introduction}
The notion of squeezing which involves reduction in the variance (uncertainty) 
of an observable below a  standard quantum limit has been studied \cite{har,rad} in the 
literature for an oscillator and a bosonic field. This notion has been extended 
to spin systems with arbitrary but sharp spin values \cite{ku,puri,wine}.
 
Quite recently, we have \cite{meh1,meh2} analysed in detail the notion of spin squeezing 
and looked into several aspects of squeezing in the case of oriented 
and non-oriented systems, and in the case of a coupled spin $s$ system composed of $2s$ 
spinor states. 
Generalizing the squeezing criterion given by Kitagawa and Ueda \cite{ku}, we have 
 made a detailed study of pure as well as mixed states.
 Kitagawa and Ueda have suggested in their paper that the occurrence of 
squeezing is a consequence of quantum spin-spin correlations that exist 
among the $2s$ spinors states which together constitute the spin $s$ state. Our 
study reveals that all oriented states are not squeezed but non-oriented 
states exhibit squeezing.  In the case of 
pure spin-1 state, our claim is that the notion of non-oriented is 
synonymous with the notion of squeezing.
 
In the light of the above studies on spin systems it becomes relevant to extend 
the ideas of squeezing to bipartite systems which do not 
possess a sharp value of spin.
 Such systems can arise due to coupling of two  systems 
with sharp angular momenta. An additional aspect that arises in such a
coupled state is whether a given state is entangled or not \cite{will}.  
An entangled state cannot be written as a product of the spin states of 
the individual systems but only as a linear combination of such products.  Further, 
it follows that the self and mutual spin-spin correlations will be present in an entangled state. 
It is therefore necessary to look for possible relationships among the three aspects viz, 
squeezing, entanglement and spin- spin correlations. 

The present paper which addresses these intrinsic notions is organized as follows.
In section 2, we look at the properties of coupled 
states and discuss the conditions to be satisfied by a coupled state to be 
a direct product state and an entangled state. In the next section, we take up the discussion on the 
 squeezing aspect for a 
coupled state which may be either entangled or not. In the case of  
uncoupled  state, the mutual spin-spin correlation will be absent and consequently 
this has to be taken into consideration while defining the squeezing 
criterion.  Taking this
 into consideration we propose a generalization of the squeezing 
criterion for a coupled state of two spin systems. A detailed presentation 
of this is given in section 3. The  dependence of squeezing on 
entanglement and on correlations is explored. 
 Section 4 deals with 
the time evolution of a coupled pure spin state in the presence of 
a spin-spin interaction. Our aim here is to show that a coupled pure spin state
undergoing such interaction develops 
squeezing as time elapses, 
even though, it may not have squeezing initially. We also look at the manner in 
which squeezing depends on spin-spin correlations. The last section is devoted 
for comments and revision.

\section{Properties of coupled states}
The $s={1 \over 2}$ states enjoy an exalted status in quantum 
theory since an arbitrary spin ${1 \over 2}$ state can always be looked 
upon either as a $\left|{1 \over 2}{1 \over 2}\right\rangle $ state or as $\left|{1 \over 2}-{1\over 2}\right\rangle$ 
state with respect to an appropriately chosen 
Cartesian frame. Thus a spinor is always oriented with respect to some $z$-axis. Following the Schwinger construction \cite{sch}, any higher spin 
$s >{1 \over 2}$ can be realized in terms of 2s spin $1\over 2$  states, 
but it cannot always be
looked upon as spin $\left|ss\right\rangle $ or $\left|s-s\right\rangle $ state with respect to any choice of the 
Cartesian frame. 
 
For $ s > \frac{1}{2}$ an oriented state is identified as an $\left|sm\right \rangle $ state  in 
an appropriately chosen Cartesian frame, with m taking any 
one of values $-s,\ldots,s$.  Such a state is 
cylindrically symmetric with respect to $\hat{z}$-axis which is also 
referred to as the axis of 
quantization. All other states are referred to as non-oriented states. 
In other words a non-oriented state cannot be looked 
upon as an $\left|sm\right\rangle $ state with respect to any frame. A 
discussion of this idea for a system with sharp angular momentum has been 
done earlier \cite{meh1,ram}. We now would like to characterize  
a coupled spin system based on these aspects. If ${\cal H}_{1}$ and ${\cal H}_{2}$ are the spin 
spaces of two systems with spins $s_{1}$ and $s_{2}$ respectively, then,  a 
coupled state of these two systems is a state belonging to ${\cal H}_{1}  
\otimes {\cal H}_{2}$ and can be expressed as 
\begin{equation}
\left|\psi _{12}\right\rangle = \sum\limits _{ij} a_{ij} \left|\phi 
_{i}\right\rangle\otimes \left|\zeta _j \right\rangle;\quad \sum\limits _{ij}\left|a_{ij}\right|^2 = 1\quad, 
\end{equation}
where $\left|\phi _{i}\right\rangle(i=- s_{1},\ldots,s_{1})$ and $\left|\zeta_j\right\rangle(j=-s_{2},\ldots,s_{2})$ 
are the angular momentum basis states of the subsystems $s_{1}$ 
and $s_{2}$ respectively. In the study of coupling of two angular 
momentum, the basis states are usually chosen with respect to some 
common axis of quantization.
For a coupled state, one possible characterization could be 
to relax this usual choice and ask whether
a state $\left|\psi _{12}\right\rangle$ is an eigen state for the four mutually commuting 
 operators 
$J^{2}_{1}$, $\vec J_{1}\cdot \hat Q_{1}$, $J^{2}_{2}$ and $\vec J_{2}\cdot \hat Q_{2}$
where $\hat Q_{1}$, $\hat Q_{2}$ refer to two arbitrary axes 
one for each system. While every state $\left|\psi _{12}\right\rangle$ is an eigen state of $J^{2}_{1}$ and $J^{2}_{2}$, 
the eigen states for other two operators form a subset of the space 
${\cal H}_{1}\otimes{\cal H}_{2}$.
   
   It is well known that in the case of two spinors the individual 
 eigen states for $J^{2}_{i}$ and $\vec J_{i}\cdot \hat Q_{i}$ 
expressed in terms of basis vectors referred to a common axis of quantization $\hat z_{0}$ of a
frame $x_{0}y_{0}z_{0}$                                                   
can be written in the form 
\begin{equation}
\left|\psi _i \right \rangle = \left(\begin {array}{c}
\cos{\textstyle{{\theta _i } \over 2}}\\[.1cm]
\sin{\textstyle{{\theta _i } \over 2}}\ e^{i\phi _i }\\[.1cm]
\end{array} 
\right)_{z_{0}},\quad 0\leq\theta _{i}\leq\pi,
\quad 0\leq\phi _{i}\leq 2\pi,\quad i = 1, 2\quad, 
\end{equation}
where $\theta _{i}$, $\phi _{i}$ are the polar angles of $\hat Q_{i}$
with respect to the common frame. From this it is clear that the  direct product state 
\begin{equation}
\fl\left|\psi^{(a)}_{_{12}}\right\rangle=\left(\begin {array}{c}
\cos{\textstyle{{\theta _1 } \over 2}}\\[.1cm]
\sin{\textstyle{{\theta _1 } \over 2}}\ e^{i\phi _1 }\\[.1cm]
\end{array} 
\right)_{ z_{0}}\otimes\left(\begin {array}{c}
\cos{\textstyle{{\theta _2 } \over 2}}\\[.1cm]
\sin{\textstyle{{\theta _2 } \over 2}}\ e^{i\phi _2 }\\[.1cm]
\end{array} 
\right)_{ z_{0}}= \left( 
\begin{array}{c}
{\cos \textstyle{\theta _{1} \over 2} \cos \textstyle{\theta _{2} \over 2}}\\[.1cm]
{\cos \textstyle{\theta _{1} \over 2}\sin \textstyle{\theta _{2} \over 2}} e^{i\phi _{2}}\\[.1cm]
{\sin \textstyle{\theta _{1} \over 2}\cos \textstyle{\theta _{2} \over 2}} e^{i\phi _{1}}\\[.1cm]
{\sin \textstyle{\theta _{1} \over 2} \sin \textstyle{\theta _{2} \over 2}} e^{i(\phi _{1}+\phi _{2})}\\
\end{array} 
\right)_{z_{0}}
\end{equation}
is an eigen state of $J^{2}_{1}$, $\vec J_{1}\cdot\hat Q_{1}$, $J^{2}_{2}$ and
$\vec J_{2}\cdot\hat Q_{2}$ satisfying (with $\hbar$=1)
\begin {equation}
 J^{2}_{1}\left|\psi^{(a)}_{_{12}}\right\rangle=\frac{1}{2}\Big(\frac{1}{2}+1\Big)\,\left|\psi^{(a)}_{_{12}}\right\rangle
\end{equation} 
\begin {equation}
\vec J_{1}\cdot \hat Q_{1}\left|\psi^{(a)}_{_{12}}\right\rangle=\frac{1}{2}\left|\psi^{(a)}_{_{12}}\right\rangle
\end{equation} 
\begin {equation}
 J^{2}_{2}\left|\psi^{(a)}_{_{12}}\right\rangle=\frac{1}{2}\Big(\frac{1}{2}+1\Big)\left|\psi^{(a)}_{_{12}}\right\rangle
\end{equation} 
\begin {equation}
\vec J_{2}\cdot\hat Q_{2}\left|\psi^{(a)}_{_{12}}\right\rangle=\frac{1}{2}\left|\psi^{(a)}_{_{12}}\right\rangle
\end{equation} 
The other three  common eigen states are 
\begin{equation}
\left|\psi^{(b)}_{_{12}}\right\rangle= \left( 
\begin{array}{c}
{\cos \textstyle{\theta _{1} \over 2} \sin \textstyle{\theta _{2} \over 2}}\\[.1cm]
-{\cos \textstyle{\theta _{1} \over 2}\cos \textstyle{\theta _{2} \over 2}} e^{i\phi _{2}}\\[.1cm]
{\sin \textstyle{\theta _{1} \over 2}\sin \textstyle{\theta _{2} \over 2}} e^{i\phi _{1}}\\[.1cm]
-{\sin \textstyle{\theta _{1} \over 2} \cos \textstyle{\theta _{2} \over 2}} e^{i(\phi _{1}+\phi _{2})}\\
\end{array}
\right)_{z_{0}}
\end{equation} 
\begin{equation}
\left|\psi^{(c)}_{_{12}}\right\rangle =\left( 
\begin{array}{c}
{\sin \textstyle{\theta _{1} \over 2} \cos \textstyle{\theta _{2} \over 2}}\\[.1cm]
{\sin \textstyle{\theta _{1} \over 2}\sin \textstyle{\theta _{2} \over 2}} e^{i\phi _{2}}\\[.1cm]
-{\cos \textstyle{\theta _{1} \over 2}\cos \textstyle{\theta _{2} \over 2}} e^{i\phi _{1}}\\[.1cm]
-{\cos \textstyle{\theta _{1} \over 2} \sin \textstyle{\theta _{2} \over 2}} e^{i(\phi _{1}+\phi _{2})}\\
\end{array}
\right)_{z_{0}}
\end{equation}
\begin{equation}
\left|\psi^{(d)}_{_{12}}\right\rangle =\left( 
\begin{array}{c}
{\sin \textstyle{\theta _{1} \over 2} \sin \textstyle{\theta _{2} \over 2}}\\[.1cm]
-{\sin \textstyle{\theta _{1} \over 2}\cos \textstyle{\theta _{2} \over 2}} e^{i\phi _{2}}\\[.1cm]
-{\cos \textstyle{\theta _{1} \over 2}\sin \textstyle{\theta _{2} \over 2}} e^{i\phi _{1}}\\[.1cm]
{\cos \textstyle{\theta _{1} \over 2} \cos \textstyle{\theta _{2} \over 2}} e^{i(\phi _{1}+\phi _{2})}\\
\end{array}
\right)_{z_{0}}.
\end{equation}
From this it is clear that every direct product state is a common eigen 
state of $J^{2}_{1}$ , $\vec J_{1}\cdot{\hat Q}_{1}$, $J^{2}_{2}$
 and $\vec J_{2}\cdot{\hat Q}_{2}$
for some ${\hat Q}_{1}$ and ${\hat Q}_{2}$ and that given one such state, 
 three  orthogonal eigen states can be  constructed, which of course
 span the direct product space ${\cal H}_{1}$ $\otimes {\cal H}_{2}$.
 Conversely it follows that the common eigen states of the above 
 four operators have to be direct product states only.
 
   In the case of a sharp spin $s$ state we have characterized 
a pure state as oriented if it happens to be an angular momentum 
state $\left|sm\right\rangle$ with respect to some quantization
axis. This property can be extended to a bipartite system of two 
spins $s_{1}$ and $s_{2}$. A pure state  $\left|\psi _{12}\right\rangle$
of a bipartite system can be regarded as oriented if 
\begin{equation}
\vec J_{1}\cdot\hat Q_{1} \ \vec J_{2}\cdot\hat Q_{2}  \left|\psi _{12}
\right\rangle=m_{1} m_{2}  \left|\psi _{12}\right\rangle
\end{equation}
for some $\hat Q_{1}$ and $\hat Q_{2}$. If $s_{1}$ and $s_{2}$ are arbitrary, then every 
direct product state $\left|\xi _{1}\right\rangle\otimes\left|\xi _{2}\right\rangle$  
 is not necessarily oriented. However if  
   $s_{1}$=$s_{2}$=$\frac{1}{2}$, every direct product  state 
 is always oriented. This follows from  the homomorphism between SU(2) and O(3).
 
  A direct product state of two spinors is thus non-entangled, oriented and   
a common eigenstate of four operators $J^{2}_{1}$ , $\vec J_{1}\cdot{\hat Q}_{1}$, $J^{2}_{2}$,
  $\vec J_{2}\cdot{\hat Q}_{2}$.
  This implies that the rest of the states in the Hilbert space ${\cal H}_{1}$ $\otimes$
${\cal H}_{2}$ do not share the above properties.  They are not only          
non-oriented but also possess entanglement. Let us now look at the 
nature of spin-spin correlations present in these coupled states. 

 The spin-spin correlations  
according to Kitagawa and Ueda are responsible for the existence of squeezing.
In their paper  \cite{ku} they suggest that every state of a spin $s$
system can be visualized as a coupled state of $2s$ spinor 
states and claim that the squeezing behaviour of the spin $s$ state 
is due to the correlations that exist among the $2s$ spinor states.
In our earlier paper \cite{meh1,meh2}, we have indeed given an explicit
method of construction of a general pure spin $s$ state in terms  of $2s$
spinor states. Based on this construction we have analysed the nature 
of a squeezed spin $s$ state and in the case of $s=1$, we have shown 
that the squeezing aspect is intimately connected with the pair-wise 
correlations defined through 
\begin{equation}
C^{ij}_{\mu\nu}(s)=\langle S^{i}_{\mu}S^{j}_{\nu}\rangle-
\langle S^{i}_{\mu}\rangle \langle S^{j}_{\nu}\rangle,\quad i,j=1,2,\ldots,.2s
\end{equation}
where $S^{i}_{\mu}$ is the $\mu^{th}$ component of the $i^{th}$ spin
$\vec S^{i}$. We wish to call these as self correlations.
 While these are absent in the case of a single spinor, 
there would be large number of such correlations for large $s$.  
 These coupled states 
of arbitrary spin $s_{1}$ and $s_{2}$ not only possess the above 
`self correlations' $C^{ij}_{\mu\nu}(s_{1})$ and $C^{ij}_{\mu\nu}(s_{2})$
but also the `mutual correlations' between $s_{1}$ and $s_{2}$.
These mutual correlations can be defined in an analogous way as 
\begin{equation}
D^{12}_{\mu\nu}=\langle S_{1\mu}S_{2\nu}\rangle-
\langle S_{1\mu}\rangle \langle S_{2\nu}\rangle
\end{equation}
where $S_{1\mu}$$(S_{2\nu})$ is the $\mu^{th}$$(\nu^{th})$ component 
of the spin vector $\vec S_{1}$ $(\vec S_{2})$. For a direct product 
state of two spinors it is easy to see that both $C^{ij}_{\mu\nu}$ and
$D^{12}_{\mu\nu}$ are zero. A direct product state with $s_{1}$ or $s_{2}$
exceeding $\frac{1}{2}$ may possess self correlations but there are no 
mutual correlations. An entangled pure state on the other hand certainly 
possess mutual correlations always, although there may or may 
not be self correlations. Our work in this paper is limited to the 
discussion of a coupled state of two spinors and we will be 
using the ideas presented here to arrive at the right criterion 
for the squeezing of such a bipartite state.

  It may be mentioned here that the most general state of two 
spins $s_{1}$ and $s_{2}$ is a mixed state which is not only entangled 
but also rich in both self and mutual correlations. In addition, 
it may possess statistical correlations owing to the 
distribution of the spins in it.
            	  	 
\section{Squeezing criterion for a general bipartite  state}
The concept of squeezing of a spin $s$ system as emphasized by Kitagawa and Ueda 
\cite{ku} is associated with quantum correlations. According to Kitagawa and Ueda 
a spin state $\left|\phi\right\rangle$ is said
to be squeezed if in that state
\begin{equation}
\Delta\big(\vec S\cdot\hat n_{\perp}\big)^{2}<\frac{|\langle\vec S\cdot \hat n\rangle|}{2}
\end{equation}
where $\hat n$ is a unit vector along $\langle \vec S\rangle$, called 
the mean spin direction and $\hat n_{\perp}$ is orthogonal to $\hat n$.
This condition clearly distinguishes a squeezed state from other
states in an intrinsic way.

The task now is to extend this to the case of a bipartite 
 system  which in general does not possess a sharp value of
 the total spin. For example, a coupled  state of two spinors
 can be  a superposition of triplet $(s=1)$ as well as the singlet
 states  $(s=0)$.  A possible choice for the criterion is to consider 
 the spin
  components $S_{1\mu}+S_{2\mu}$ of the total spin $\vec S=\vec S_{1}+
  \vec S_{2}$  with respect to a 
  common frame $x_{0}y_{0}z_{0}$ and define the criterion exactly as 
  in (Eq.14)
  with the understanding that $\hat n$ denotes the direction of 
  $\langle \vec S_{1}+\vec S_{2}\rangle$ and $\vec S .\hat n$,
  $\vec S .\hat n_{\perp}$        
  denote the components of $\vec S=\vec S_{1}+\vec S_{2}$ along and perpendicular to    
$\hat n$  respectively .
  We have looked into this choice which leads to a conclusion that
  certain direct product states such as
\begin{equation}
\left|\psi\right\rangle=
\left( 
\begin{array}{c}
{\frac{\sqrt 3}{2}}\\[.1cm]
 {\frac{1}{2}}\\
\end{array} 
\right)\otimes\left( 
\begin{array}{c}
{\frac{\sqrt 3}{2}}\\[.1cm]
 {\frac{-1}{2}}\\
\end{array} 
\right)=
\left( 
\begin{array}{c}
{\frac{3}{4}}\\[.1cm]
-{\frac{\sqrt{3}}{4}}\\[.1cm]
{\frac{\sqrt{3}}{4}}\\[.1cm]
 -{\frac{1}{4}}\\
\end{array} 
\right),
\end{equation}
will possess  squeezing if we consider such a choice.
This is undesirable since  the two subsystems  may be totally independent, 
non squeezed and non interacting and if we 
formally define direct product states of the two,
 these would be squeezed if we employ the criterion suggested 
 above.
   
 We would therefore like to search for  an appropriate 
criterion for squeezing of bipartite states taking the above aspect
into consideration.
As an aid in this direction, it may be noted here that 
in problems aimed at determining correlations in a 
bipartite system, when the subsystems are physically separated, two observers
make measurements in frames of their own choice.
This freedom of choice of frame, in the context of entanglement, arises from the property that the entanglement
of a bipartite system is invariant under local rotation of frames (local unitary
transformations on individual states ) describing the subsystems.
It is therefore necessary to allow for this freedom of choice of local frames for discussing squeezing 
and this is done as follows.

 Suppose a bipartite state $\left|\psi _{12}\right\rangle$ (Eq.1) of two spins is 
specified with respect to some frame $x_{0} y_{0} z_{0}$ (say).
  Suppose $[\hat n_{1},  \hat n_{1\perp},  \hat n_{1\perp^{\prime}}=\hat n_{1}\times\hat n_{1\perp}]$ and
$[\hat n_{2},  \hat n_{2\perp},  \hat n_{2\perp^{\prime}}=\hat n_{2}\times\hat n_{2\perp}]$
denote two sets of mutually orthogonal directions. Now it is easy to see  
that the components of spin operators $\vec S_{1}$ and $\vec S_{2}$
with respect to these triplets satisfy the usual angular momentum 
commutation relations,
\begin{equation}
\Big[\vec S_{1}.\hat n_{1\perp}+\vec S_{2}.\hat n_{2\perp},  \vec S_{1}.\hat n_{1\perp ^{\prime}} +\vec S_{2}.\hat n_{2\perp ^{\prime}}\Big]=
i \Big(\vec S_{1}.\hat n_{1}+\vec S_{2}.\hat n_{2}\Big)
\end{equation}
and the uncertainty relationship for these operators takes the form
\begin{equation}
\fl\Delta\Big(\vec S_{1}.\hat n_{1\perp}+\vec S_{2}.\hat n_{2\perp}\Big)^{2}
\quad \Delta\Big(\vec S_{1}.\hat n_{1\perp^{\prime}}+\vec S_{2}.\hat n_{2\perp^{\prime}}\Big)^{2}\geq
\frac{\Big(\langle \vec S_{1} .\hat n_{1}\rangle+ \langle \vec S_{2} .\hat n_{2}\rangle\Big)^{2}} {4},
\end{equation}
where,
\begin{equation}
\fl\Delta\Big(\vec S_{1}\cdot\hat a+\vec S_{2}\cdot\hat b\Big)^{2}=
\langle\psi _{12}\left|(\vec S_{1}\cdot\hat a+\vec S_{2}\cdot
\hat b)^{2}\right|\psi _{12}\rangle-\langle\psi _{12}\left|(\vec S_{1}\cdot\hat a+\vec S_{2}\cdot\hat b)
\right|\psi _{12}\rangle ^{2}
\end{equation}

Suppose now $\hat n_{1}$ and
 $\hat n_{2}$ denote the mean spin directions of the individual 
 spinor states, defined through 
 \begin{equation}
 \hat n_{i}=\frac{\langle\psi _{12}|\vec S_{i}|\psi _{12}\rangle}{\left|\langle\psi _{12}|\vec S_{i}|\psi _{12}\rangle\right|},\quad i=1,2
 \end{equation}
and $\hat n_{i\perp }$ are directions such that
\begin{equation}
\hat n_{i\perp}\cdot\hat n_{i}=0,\quad i=1,2
\end{equation}
These directions $(\hat n_{i}, \hat n_{i\perp}, \hat n_{i\perp^{\prime}})$ 
define the individual Lakin frames \cite{meh1} as we have
 \begin{equation}
 \bigl\langle\vec S_{i}\cdot\hat n_{i_\perp}\bigr\rangle=0,\quad i=1,2 .
 \end{equation}
The criterion we adopt is as follows. Given a bipartite state 
there are two mean spin directions $\hat n_{1}$ and  $\hat n_{2}$ defined 
through equation (19). A bipartite state with mean spin directions
$\hat n_{1}$, $\hat n_{2}$ is said to be squeezed in a perpendicular
component $\vec S_{1}\cdot\hat n_{1\perp}+\vec S_{2}\cdot\hat n_{2\perp}$, 
if the variance of this operator in the given state is less than 
half the sum of the absolute values of the expectation values of the spin vectors 
in that state.

Expressed mathematically, the criterion becomes  		 
 \begin{equation}
 \Delta\big(\vec S_{1}\cdot\hat n_{1_\perp}+\vec S_{2}\cdot\hat n_{2_\perp}\big)^{2}<
 \frac{|\langle\vec S_{1}\rangle|+|\langle\vec S_{2}\rangle|}{2}
\end{equation}		 
This can be further written in the form
\begin{eqnarray}		 
\fl\Delta\big(\vec S_{1}\cdot\hat n_{1_\perp}\big)^{2}	+\Delta\big(\vec S_{2}\cdot\hat n_{2_\perp}\big)^{2}+
2\langle \vec S_{1}\cdot\hat n_{1_\perp}\otimes\vec S_{2}\cdot\hat n_{2_\perp}\rangle
<\frac{|\langle\vec S_{1}\cdot\hat n_{1}\rangle|+|\langle\vec S_{2}\cdot\hat n_{2}\rangle|}{2}.
\end{eqnarray}
Before we provide some justification for this criterion, it must be noted that it is 
in an invariant form so that given  any frame, the criterion
can be expressed in terms of the spin components referred to that frame, once the 
directions $\hat n_{i}$, $\hat n_{i\perp}$ are determined in that frame.
Instead, one can also transform the frame by appropriate rotations 
and obtain the individual Lakin frames. Suppose the so obtained frames
$x_{1}y_{1}z_{1}$ and $x_{2}y_{2}z_{2}$ are named such that 
$\hat z_{i}=\hat n_{i}$, $\hat x_{i}=\hat n_{i\perp}$ and
$\hat y_{i}=\hat n_{i\perp^{\prime}}$,
 the criterion takes the forms 
 \begin{eqnarray} 
 \Delta\big(S_{1x_{1}}\big)^{2}+\Delta\big(S_{2x_{2}}\big)^{2}+
 2\langle S_{1x_{1}}\otimes S_{2x_{2}}\rangle<\frac{|\langle\vec S_{1z_{1}}\rangle|+|\langle\vec S_{2z_{2}}\rangle|}{2}
  \end{eqnarray}
  \begin{equation} 
  \Delta\big(S_{1y_{1}}\big)^{2}+\Delta\big(S_{2y_{2}}\big)^{2}+
  2\langle S_{1y_{1}}\otimes S_{2y_{2}}\rangle<\frac{|\langle\vec S_{1z_{1}}\rangle|+|\langle\vec S_{2z_{2}}\rangle|}{2}
 \end{equation}
  
The given state would therefore be squeezed in the components
$S_{1x_{1}}+S_{2x_{2}}$ or $S_{1y_{1}}+S_{2y_{2}}$ 
if equation (24) or (25) is satisfied.
For a bipartite system of two spinors, the criterion reduces to 
a simpler form since, 
\begin{eqnarray}
\Delta  S^{2}_{1{x_{1}}} =\Delta  S^{2}_{2{x_{2}}}=\Delta  S^{2}_{1{y_{1}}}=\Delta  S^{2}_{2{y_{2}}}=\frac{1}{4}
\end{eqnarray}
always and therefore the squeezing criterion along the individual x and y-axes
reduces to
\begin{equation}
\langle S_{1{x_{1}}}\otimes S_{2{x_{2}}}\rangle<\frac{|\langle\vec S_{1{z_{1}}}\rangle|+|\langle\vec S_{2{z_{2}}}\rangle|-1}{4}
\end{equation}
  and in y-direction
\begin{equation}
\langle S_{1{y_{1}}}\otimes S_{2{y_{2}}}\rangle<\frac{|\langle\vec S_{1{z_{1}}}\rangle|+|\langle\vec S_{2{z_{2}}}\rangle|-1}{4}
\end{equation}
  
 We wish  to now apply this criterion to bipartite 
states of interest and see whether they are squeezed or not. To begin with 
let us consider a bipartite state which is a direct product state
of two states with the first being $\left|\frac{1}{2}\frac{1}{2}\right\rangle$
 with respect to $\hat z_{1}$ while the second  
 $\left|\frac{1}{2}\frac{1}{2}\right\rangle$ with respect to $\hat z_{2}$.
When  expressed in terms of a common frame $x_{0}y_{0}z_{0}$
and in individual Lakin frames it has the structure 
\begin{equation}
\left|\psi _{12}\right\rangle=\left|\xi _{1}\right\rangle \otimes\left|\xi _{2}\right\rangle
= \left( 
\begin{array}{c}
{\cos ^2\textstyle{\theta \over 2}}\\[.1cm]
-{\sin \textstyle{\theta \over 2}\cos \textstyle{\theta \over 2}}\\[.1cm]
{\cos \textstyle{\theta \over 2}\sin \textstyle{\theta \over 2}}\\[.1cm]
 -{\sin ^2\textstyle{\theta \over 2}}\\
\end{array} 
\right)_{z_{0}}= \left(
\begin{array}{c}
{1}\\[.1cm]
0\\[.1cm]
\end{array} 
\right)_{z_{1}}\otimes \left(
\begin{array}{c}
{1}\\[.1cm]
0\\[.1cm]
\end{array} 
\right)_{z_{2}}
\end{equation}
Since
\begin{equation}
 \Delta\big(\vec S_{1}\cdot\hat n_{1_\perp}\big)^{2}
 =\Delta\big(\vec S_{2}\cdot\hat n_{2_\perp}\big)^{2}=\frac{1}{4},
 \end{equation}
 \begin{equation}
 \langle \vec S_{1}\cdot\hat n_{1_\perp}\otimes \vec S_{2}\cdot\hat n_{2_\perp}\rangle=
 \langle \vec S_{1}\cdot\hat n_{1_\perp}\rangle
 \otimes \langle \vec S_{2}\cdot\hat n_{2_\perp}\rangle=0,
 \end{equation}
\begin{equation}
 \left|\langle \vec S_{1}\cdot\hat z_{1}\rangle\right|=
  \left|\langle \vec S_{2}\cdot\hat z_{2}\rangle\right|=\frac{1}{2},
\end{equation}
 the criterion in the form (23) or in the forms (24), (25)
  is not satisfied at all and hence a direct product 
 state of two spinors is never squeezed.  This is in perfect agreement with 
 Kitagawa and Ueda \cite{ku} in that the squeezing arises due to correlations and 
 a direct product state which has neither self nor mutual  correlations is therefore not squeezed. 
 
\subsection{Squeezing of entangled pure state}
It is clear from the previous discussion that entanglement is necessary
in the case of a two spinor bipartite state 
for squeezing to occur. However it is to be seen whether entanglement is sufficient 
also. We therefore begin this study by considering 
a general pure state which is entangled. Such a state can be expressed 
with respect to a basis $\left|m_{1}m_{2}\right\rangle _{z_{0}}$
referred to a common frame $x_{0}y_{0}z_{0}$ in the form  
\begin{equation} 
\left| {\psi _{12} } 
\right\rangle = \left( {\begin{array}{l}
 a_{11} \\ 
 a_{12} \\ 
 a_{21} \\ 
 a_{22} \\ 
 \end{array}} \right)\quad ;\quad\sum _{ij}\left|a_{ij}\right|^{2}=1
\end{equation}
 where of course $a_{11} a_{22} \neq a_{12} a_{21} $. This 
 latter condition ensures \cite{will}  that $\left| {\psi _{12} }\right\rangle$ is entangled.
 Since in the general case the frame $x_{0}y_{0}z_{0}$ may not be a Lakin frame
 for either spinor, we consider the rotation
 \begin{equation}
R_{12} = R_1 \left( {\phi _1, \theta _1, 0} \right) \otimes R_2 \left( {\phi _2, 
\theta _2, 0} \right)
\end{equation}
on this state so that
 \begin{equation}
\left|\psi _{12}\right\rangle\longrightarrow\left|\psi ^{\prime} _{12}\right\rangle=
 R_{12}\left| {\psi _{12} }\right \rangle = 
\left( {\begin{array}{l}
 c_{11} \\ 
 c_{12} \\ 
 c_{21} \\ 
 c_{22} \\ 
 \end{array}} \right)
\end{equation}
where $c_{11}$ can be taken to be a non negative real number (by using 
the freedom of choice of the over all phase). The individual rotations
on the coordinate system $x_{0}y_{0}z_{0}$ take it to the respective
Lakin frames $x_{1}y_{1}z_{1}$ and $x_{2}y_{2}z_{2}$ if the Euler angles of rotation are chosen to satisfy 
\begin{equation}
\tan\phi _{i}= \frac{\langle {S_{iy } } \rangle 
}{\langle {S_{ix } } \rangle },\quad i=1,2
\end{equation}
\begin{equation}
\tan \theta _{i} = \frac{\left(\langle S_{iy}\rangle^2
+\langle S_{ix }\rangle^2\right)^{1\over 2}}{\langle S_{iz}\rangle},\quad i=1,2
\end{equation}
With these transformations we now obtain
\begin{equation}
\langle {S_{1x_{_{1}} } } \rangle=\langle {R_{12}S_{1x}R_{12}^{\dagger}}
\rangle   
={1\over 2}\left({c_{11} c_{21} + c_{12}^{\star} 
c_{22} + c_{22}^ {\star} c_{12} + c_{21}^ {\star} c_{11} } \right) = 0
\end{equation}
\begin{equation}
\langle {S_{2x_{_{2}} } } \rangle=\langle
{R_{21}S_{2x}R_{21}^{\dagger}} \rangle   
 = {1 \over 2}\left( {c_{11} c_{12} + c_{12}^{\star} 
c_{11} + c_{21}^{\star} c_{22} + c_{22}^{\star} c_{21} } \right) = 0
\end{equation}
\begin{equation}
\langle {S_{1y_{_{1}} } } \rangle=\langle
{R_{12}S_{1y}R_{12}^{\dagger}} \rangle   
 ={i\over 2}\left( { - c_{11}  c_{21} - c_{12}^{\star}
c_{22}+c_{21}^{\star} c_{11} + c_{22}^{\star} c_{12} } \right) = 0
\end{equation}
\begin{equation}
\langle {S_{2y_{_{2}} } } \rangle=\langle
{R_{21}S_{2y}R_{21}^{\dagger}} \rangle   
= {i \over 2}\left( { - c_{11}  c_{12} + c_{12}^{\star} 
c_{11} - c_{21}^{\star} c_{22} + c_{22}^{\star} c_{21} } \right) = 0
\end{equation}
\begin{equation}
\langle {S_{1z_{_{1}} } } \rangle=\langle
{R_{12}S_{1z}R_{12}^{\dagger}} \rangle  
={1 \over 2}( c_{11}^2 + 
\left|{c_{12} } \right|^2 - \left| {c_{21} } \right|^2 - \left| {c_{22} } 
\right|^2)
\end{equation}
\begin{equation}
\langle {S_{2z_{_{2}} }} \rangle=\langle
{R_{21}S_{2z}R_{21}^{\dagger}} \rangle  
={1 \over 2}( c_{11}^2 - 
\left|{c_{12} } \right|^2 + \left| {c_{21} } \right|^2 - \left| {c_{22} } 
\right|^2)
\end{equation}
A glance at the squeezing criterion implies that for the state 
to exhibit squeezing, first of all, $\langle S_{1z_{1}}\rangle$, 
 $\langle S_{2z_{2}}\rangle \neq 0$. Further the first 
 four equations referring to Lakin frame yield 
\begin{equation}
c_{11}  c_{21} = - c_{12}^{\star}  c_{22} 
\end{equation}
\begin{equation}
c_{11}  c_{12} = - c_{21}^{\star} c_{22} 
\end{equation}
which lead to 
\begin{equation}
c_{11}c_{22}(\left| c_{21} \right|^2-\left| c_{12} \right|^2)\quad=\quad 0
\end{equation}
and
\begin{equation}
c_{12}^{\star}c_{21}( c_{11}^2-\left| c_{22} \right|^2)\quad=\quad 0
\end{equation}
There arise several  cases satisfying these conditions
\begin{equation}
{\rm case\, 1:}\quad c_{ij} =\delta _{ii_{0}}\delta _{jj_{0}}\quad
{\rm for\ fixed}\quad  i_{0},j_{0}
\end{equation}
\begin{equation} 
{\rm case\, 2:}\quad\left| {c_{21} } \right| = \left| {c_{12} } \right| \neq 0,\quad 
\left| {c_{11} } \right| = \left| {c_{22} } \right|,\quad\phi _{12}=\pi+\phi _{22}-\phi _{21} 
\end{equation}
\begin{equation}
{\rm case\, 3:}\quad c_{11}=c_{22}=0,\quad c_{12},c_{21}\neq 0
\end{equation}
\begin{equation} 
{\rm case\, 4:}\quad\left| {c_{12} } \right| = \left| {c_{21} } \right| = 0  
\end{equation}
The first case refers to direct product states. The  second case  implies
$\langle S_{1z_{1}}\rangle$ = $\langle S_{2z_{2}}\rangle$=0
and hence although it is entangled, it is not squeezed. 
 This state is actually a singlet state with total spin $s=0$ 
and $\left|\langle S_{1z_{1}}\rangle\right|$+$\left|\langle S_{2z_{2}}\rangle\right|=0$.  

 It is this latter value which makes a singlet state not squeezed 
although there is entanglement. This state therefore implies that the 
entanglement is necessary but not sufficient.  
    Consider now the state in case (3), which has the form
\begin{equation}
\left| \xi \right\rangle = 
\left( {\begin{array}{c}
 0\\ 
 c_{12} \\ 
 c_{21}\\ 
 0 \\ 
 \end{array}} \right)\quad .
\end{equation}
Under the rotation $R_{1}(0,\frac{\pi}{2},0)\otimes I$, which is a local rotation, this state changes to
\begin{equation}
\left| \xi^{\prime} \right\rangle =(R_{1}\otimes I)\left| \xi \right\rangle=
\left( {\begin{array}{cccc}
 0& 0& 1& 0\\ 
 0& 0& 0& 1\\ 
 -1& 0& 0& 0\\
 0&-1& 0& 0\\ 
 \end{array}} \right)
 \left( {\begin{array}{cccc}
  0\\ 
 c_{12}\\ 
 c_{21}\\
 0\\ 
 \end{array}} \right)=
\left( {\begin{array}{cccc}
  c_{21}\\ 
 0\\ 
 0\\
 -c_{12}\\ 
 \end{array}} \right)
\end{equation}
which shows that it belongs to case (4). It is therefore
enough to consider only the states belonging to case (4). Here too, 
if we use the degree of freedom for the over all phase, the normalization 
condition and the freedom of rotation about the respective $z_{i}$ axes, 
the state can be reduced to the simple form with its elements parametrized as 
\begin{equation}
\left| \chi \right\rangle = \left( {\begin{array}{c}
 \cos \frac{\theta}{2}\\ 
 0 \\ 
 0 \\ 
 \sin \frac{\theta}{2} \\ 
 \end{array}} \right),\quad 0<\theta<\pi.
\end{equation}
This is the simplest matrix form of an entangled state with non zero mean values 
for its individual spin vectors.  The relevant quantities needed to determine 
the squeezing behaviour are
\[
\Delta S_{1x_1 }^2 = \Delta S_{2x_2 }^2 = \Delta S_{1y_1 }^2 = \Delta 
S_{2y_2 }^2 = \frac{1}{4}
\]
\[
\langle {S_{1x_1 } \otimes S{ }_{2x_2 }} \rangle = \textstyle{1 
\over 4}\sin \theta  = - \langle {S_{1y_1 } \otimes S_{2y_2 } 
} \rangle 
\]
\[
\left|\langle S_{1_{z_{1}}} \rangle\right|=\left|\langle S_{2_{z_{2}}} \rangle\right| =\frac{1}{2}\left| \cos \theta\right|\quad . 
\] 
Substituting  these in the squeezing condition given by equations (23) or (24) we obtain   
\begin{equation}
1+\sin\theta<\left|\cos\theta\right|\quad {\rm for}\, S_{1x_1}+S_{2x_2}
\end{equation}
and
\begin{equation}
1-\sin\theta<\left|\cos\theta\right|\quad {\rm for}\, S_{1y_1}+S_{2y_2}
\end{equation}

These conditions are certainly satisfied for a wide range of 
$\theta$ which indicates that a wide 
variety of states of the form (Eq.54) with different values of $\theta$
exhibit squeezing.
The  variation of squeezing with respect to
 $\theta$ is shown in the graph below  figure (1) where we have 
 plotted the difference between the right hand and left hand sides of equation (55) and (56)
\begin{equation}
Q_{x}=\left|\cos\theta\right|-\sin\theta-1
\end{equation}
\begin{equation}
Q_{y}=\left|\cos\theta\right|+\sin\theta-1
\end{equation}  
as a function of  $\theta$.
Positive (negative) values of $Q_{x}$, $Q_{y}$ show the presence (absence) of squeezing. 
It may be noted from the figure (1) that $\theta=90^{\circ}$, corresponds to the singlet state 
referred to in case(2).
    
     At this stage it is relevant to see how the correlations account for squeezing. 
As mentioned earlier, a bipartite system of two spinors has no self correlations. 
The mutual correlations which exist between the two spins are,  
\begin{equation}
D_{\mu _1\nu _2 }^{12} = \langle {S_{1\mu _1 } S_{2\nu _2 } } 
\rangle - \langle {S_{1\mu _1 } } \rangle \langle 
{S_{2\nu _2 } } \rangle \quad\quad\mu ,\nu \quad = x,y,z. 
\end{equation}
If these are zero, then the state has no mutual spin-spin correlation.
All direct product states fall into this category.  For the squeezed states defined 
by (Eq.54) these correlations with respect to the Lakin frames turn out 
to be
\begin{equation}
D_{x_1x_2}^{12} = - D_{y_{1}y_{2}}^{12} = \frac{1}{4}\sin \theta  
\end{equation}
\begin{equation}
D_{z_1 z_2 }^{12} =\frac{1}{4}\sin^{2}\theta
\end{equation}
\begin{equation}
D_{x_{1}y_{2}}^{12} =D_{x_{1}z_{2}}^{12}=D_{y_{1}z_{2}}^{12} =0  
\end{equation}

The graphs shown in figure (2) in which we have plotted both the 
squeezing and mutual correlation functions reveal that squeezing is absent
whenever there are no mutual correlations and when the mutual 
correlations assume their extreme values. Indeed these extreme  values correspond 
to a singlet state which confirms that it is a maximally entangled state. However,
it has both $\langle \vec S_{1}\rangle =\langle \vec S_{2}\rangle\,=0$ 
due to which it does not exhibit squeezing.
 
\begin{center}
\begin{figure}[H]
\vspace*{-1in}
\includegraphics[width=\linewidth]{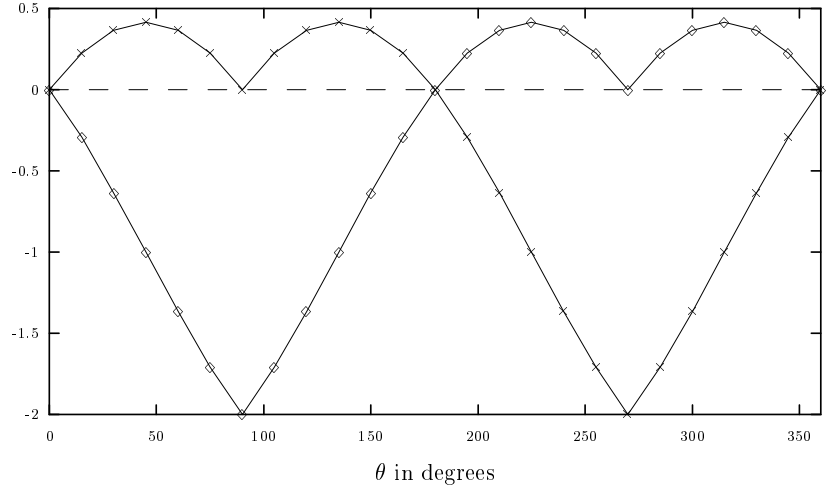}\\
\vspace*{-5in}
\caption{Variation of squeezing  $Q_{x}(\diamond)$ and 
 $Q_{y}(\times)$ with respect to $\theta$.} 
\end{figure}
\end{center}

\begin{figure}[H]
\begin{center}
\vspace*{-1in}
\includegraphics[width=\linewidth]{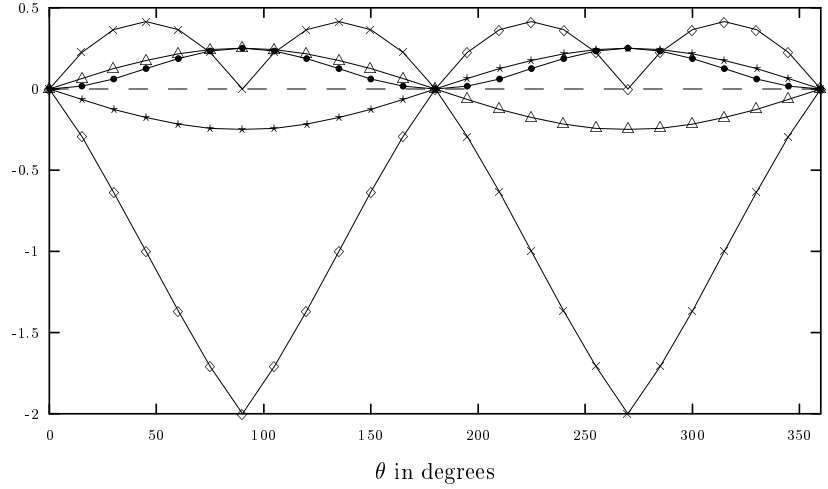}\\
\vspace*{-5in}
\caption{Variation of spin-spin correlations $D^{12}_{xx}(\triangle)$,
$D^{12}_{yy}(\star)$ and $D^{12}_{zz}(\circ)$, The plot also show  
squeezing in $Q_{x}(\diamond)$ and 
 $Q_{y}(\times)$ with respect to $\theta$.}  
\end{center}
\end{figure}

\section{Generation of squeezing}
We have so far discussed the nature of spin squeezing for a general coupled 
pure state in terms of a new criterion which is  in some sense a generalization 
of the earlier criterion. It is of interest to know whether squeezing can be 
generated by subjecting a spin system to external interactions. Indeed such 
an attempt has been done by Kitagawa and Ueda  for sharp
spin state \cite{ku} earlier where they 
consider a Hamiltonian quadratic in the spin operators and show that the 
state generated during evolution under such a Hamiltonian possesses squeezing. We follow their 
prescription to see whether squeezing can be generated in the case of a 
coupled state too. It is clear that an Hamiltonian linear in the spin 
variables can only lead to a rotation of the state or of the coordinate system. Such an action does not 
introduce any squeezing. It is therefore necessary that the Hamiltonian 
should be at least quadratic in the spin variables.  We therefore consider below a spin-spin 
interaction Hamiltonian 
\begin{equation}
H = i \xi\left[ {S_{1 + } S_{2 + } - S_{1 - } S_{2 - } } \right] 
\end{equation}
where,
\begin{equation}
 S_{i \pm } = S_{ix} \pm i S_{i y},\quad i=1,2
 \end{equation}
  and $\xi$ is any real number. If the initial coupled 
state $\left| \psi \right\rangle $ of these two spinors is chosen to be a 
direct product state $\left| \psi \right\rangle $ as in (Eq.29) where the 
basis vectors and the Hamiltonian  are referred to the frame x$_{0}$y$_{0}$z$_{0}$, 
the evolution in time given by
\begin{equation}
\left| {\psi (t)} \right\rangle =
 \exp{(-iHt)}\left| \psi\right\rangle=\exp{[\xi t(S_{1{+}}S_{2{+}}-S_{1{-}}S_{2{-}})]}\left| \psi \right\rangle 
\end{equation}
leads to the explicit form
\begin{equation}
\left| {\psi (t)} \right\rangle = \left( {\begin{array}{c}
 \cos \tau \cos ^2\textstyle{\theta \over 2} - \sin \tau \sin 
^2\textstyle{\theta \over 2} \\[.1cm] 
 - \cos \textstyle{\theta \over 2}\sin \textstyle{\theta \over 2} \\[.1cm] 
 \cos \textstyle{\theta \over 2}\sin \textstyle{\theta \over 2} \\[.1cm] 
  - \sin \tau \cos ^2\textstyle{\theta \over 2}-\cos \tau \sin 
^2\textstyle{\theta \over 2} \\ 
\end{array}} \right);\ \tau=\xi t.
\end{equation}
This state is in general entangled. As in any general case, we find here 
also that the frame x$_{0}y_{0}z_{0}$ is not the common  Lakin frame for the 
two spinors, since  
\begin{equation}
\langle {S_{1{x_0 } } } \rangle = \frac{\sin \theta }{2}\left[ 
{\cos \tau + \sin \tau \cos \theta } \right] = - \langle {S_{2{x_0 } } 
} \rangle 
\end{equation}
\begin{equation}
\langle {S_{1{y_0 } } } \rangle = \langle {S_{2{y_0 } } } 
\rangle = 0
\end{equation}
\begin{equation}
\langle {S_{1{z_0 } } } \rangle = \langle {S_{2{z_0 } } } 
\rangle = \frac{1}{2}\left[ {\cos 2\tau \cos \theta - \frac{1}{2}\sin 
2\tau \sin ^2\theta } \right].
\end{equation}
It is easier to analyse the squeezing behaviour if we go over to the individual Lakin 
frames x$_{1}$y$_{1}$z$_{1}$ and x$_{2}$ y$_{2}$z$_{2}$ via  the rotations through
\begin{equation}
 \alpha _i =\tan^{-1}\Big(\frac{\langle {S_{i{x_0 } } }
\rangle}{\langle {S_{i{z_0 } } } \rangle}\Big) \quad ;\quad i = 1,2
\end{equation}
of x$_{0}$y$_{0}$z$_{0}$ about y$_{0}$-axis.

In these Lakin frames the expectation values of the various spin operators are 
given by 
\begin{equation}
\fl\langle {S_{ix_{i}}}\rangle = \langle {S_{ix_{0}}} 
\rangle \cos \alpha _i - \langle {S_{iz_{0}}} \rangle 
\sin \alpha _i = 0
\end{equation}
\begin{equation}
\fl\langle {S_{iy_{i}}}\rangle =\langle {S_{iy_{0}}}
\rangle= 0
\end{equation}
\begin{equation}
\fl\langle {S_{iz_{i}}}\rangle = \langle {S_{ix_{0}}} 
\rangle \sin \alpha _i + \langle {S_{iz_{0}}} \rangle 
\cos \alpha _i 
\end{equation}
\begin{equation}
\fl\Delta S_{1x_1 }^2 + \Delta S_{2x_2 }^2 + 2\langle {S_{1x_1 } 
\otimes S_{2x_2 } } \rangle \\ 
=\frac{1}{2} (1- A \cos ^2\alpha _{1} -B \sin 2\alpha _{1} \sin \theta-
\cos ^2\theta \sin ^2\alpha _{1})  
\end{equation}
\begin{equation}
\fl\Delta S_{1y_1 }^2 + \Delta S_{2y_2 }^2 + 2\langle {S_{1y_1 } \otimes 
S_{2y_2 } } \rangle = \frac{1}{2} + \frac{1}{2}\left[ {\sin 2\tau \cos 
\theta - \sin ^2\theta \sin ^2\tau } \right]
\end{equation}
\begin{equation}
\fl\left|\langle S_{1z_1 }\rangle\right| +\left|\langle S_{2z_2 }  \rangle \right|= ( {\sin 
^2\theta \left( {\cos \tau + \sin \tau \cos \theta } \right)^2 + \left( 
{\cos \theta \cos 2\tau - \textstyle{1 \over 2}\sin 2\tau \sin ^2\theta } 
\right)^2} )^{1\over 2}
\end{equation}
where $A=(\cos \theta \sin 2\tau + \sin ^2\theta \cos ^2\tau )$ and
 $B= \sin \tau - \cos \theta \cos \tau$. The quantities $Q_{x}$, $Q_{y}$
 defined earlier now become function of time and are given by 

\begin{eqnarray}
\fl Q_{x}(t) =\Big ( {\sin 
^2\theta \left( {\cos \tau + \sin \tau \cos \theta } \right)^2 + \left( 
{\cos \theta \cos 2\tau - \textstyle{1 \over 2}\sin 2\tau \sin ^2\theta } 
\right)^2}\Big )^{1\over 2}\nonumber \\\lo-(1- A \cos ^2\alpha _{1} -
B \sin 2\alpha _{1} \sin \theta-
\cos ^2\theta \sin ^2\alpha _{1})   
\end{eqnarray}
\begin{eqnarray}
\fl Q_{y}(t) =\Big ( {\sin 
^2\theta \left( {\cos \tau + \sin \tau \cos \theta } \right)^2 + \left( 
{\cos \theta \cos 2\tau - \textstyle{1 \over 2}\sin 2\tau \sin ^2\theta } 
\right)^2}\Big )^{1\over 2}\nonumber \\\lo -(1 +  {\sin 2\tau \cos 
\theta - \sin ^2\theta \sin ^2\tau } ) 
\end{eqnarray}
We infer that the state is squeezed if either $Q_x \left( t \right)$ or $Q_y 
\left( t \right)$ is positive. The graphs of $Q_x \left( t \right)$ and $Q_y 
\left( t \right)$ plotted below in the figure (3) show that the squeezing is 
observed for a wide range of values of $\theta $ and $\tau $, except at 
certain points. These points are at $\tau=90^{\circ}$, $\theta=\frac {n\pi}{2}$,
$n = 0,1,2,\ldots$ and for $\tau=45^{\circ}$, $\theta =0$ etc, and
 correspond to either direct product or singlet states.

We now look at the mutual correlations that exist between the two spinors at 
various instants of time evolution. These are explicitly given by
\begin{eqnarray}
\fl D_{x_1 x_2 }^{12} = - \frac{1}{4}(A\cos^{2}\alpha _{1}+B\sin 2\alpha _{1}\sin\theta
  +\cos^{2}\theta\sin^{2}\alpha _{1})
\\
\fl D_{y_1 y_2 }^{12} = \frac{1}{4}\left( {\sin 2\tau \cos \theta - \sin 
^2\theta \sin ^2\tau } \right)
\\
\fl D_{z_1 z_2 }^{12} = \frac{1}{4} \big[A\sin^{2}\alpha _{1}-B \sin\theta
 \sin 2\alpha _{1} +\cos^{2}\theta\cos^{2}\alpha _{1}
 -(\sin^2\theta \left(\cos \tau + \sin \tau \cos \theta\right)^2\nonumber\\ +4 \left( 
\cos \theta \cos 2\tau - \textstyle{1 \over 2}\sin 2\tau \sin ^2\theta  
\right)^2 )\big]
\\[.2cm]
\fl D_{x_1 z_2 }^{12} = -\frac{\sin 2\alpha _{1} }{8}\left(B^{2}-\sin^{2}\theta \right)- 
\frac{B\sin\theta\cos 2\alpha _{1}}{4} 
\\
\fl D_{x_1 y_2 }^{12} = D_{y_1 z_2 }^{12} = 0.
\end{eqnarray} 
Plotting these correlations together with squeezing functions,
we observe from the figure (4) below that whenever there is 
squeezing the state necessarily possesses spin-spin correlations. 
The graphs lead to similar conclusions arrived earlier in the 
discussion of the general case.

\begin{figure}[H]
\begin{center}
\vspace*{-1in}
\includegraphics[width=\linewidth]{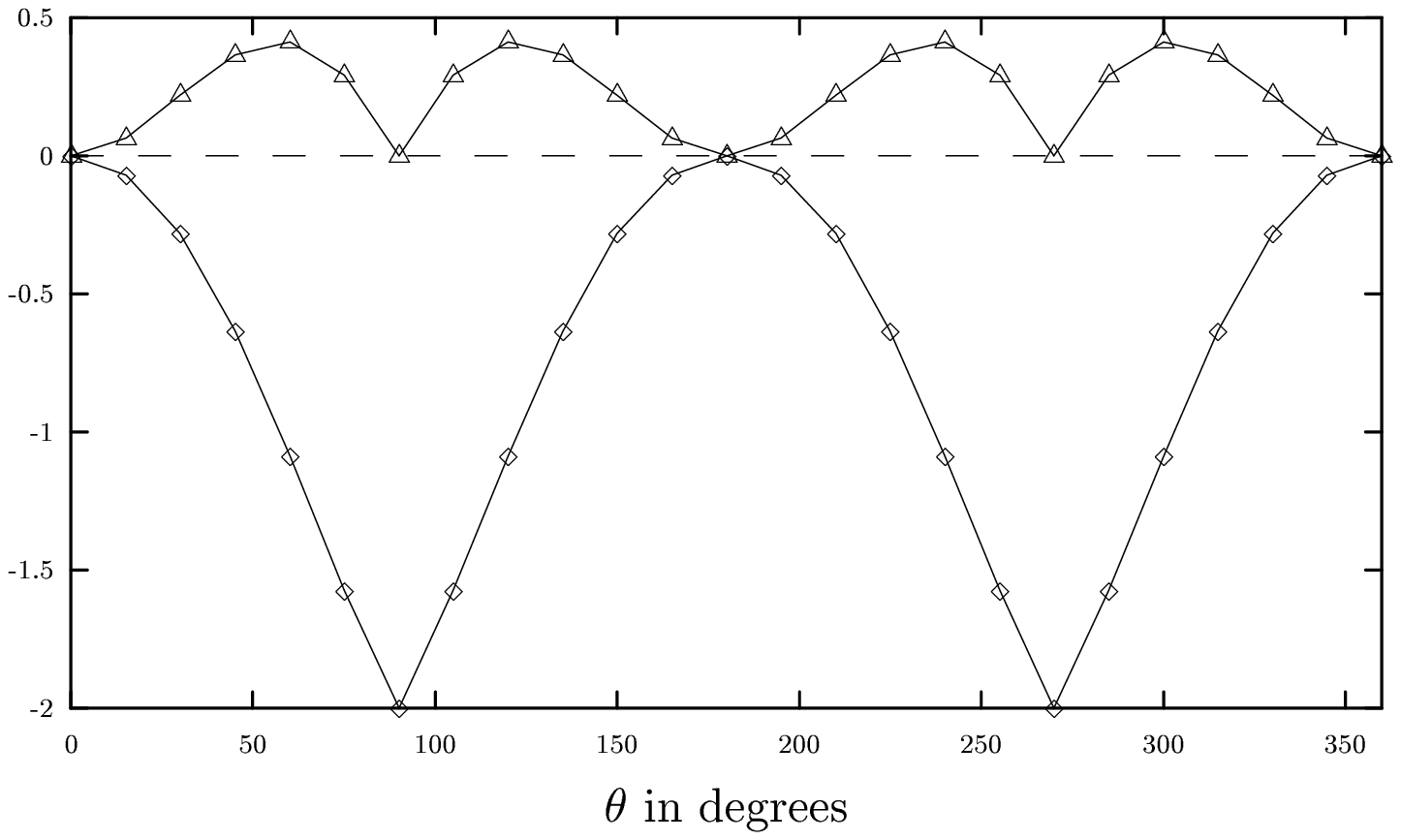}\\
\vspace*{-5.5in}
\includegraphics[width=\linewidth]{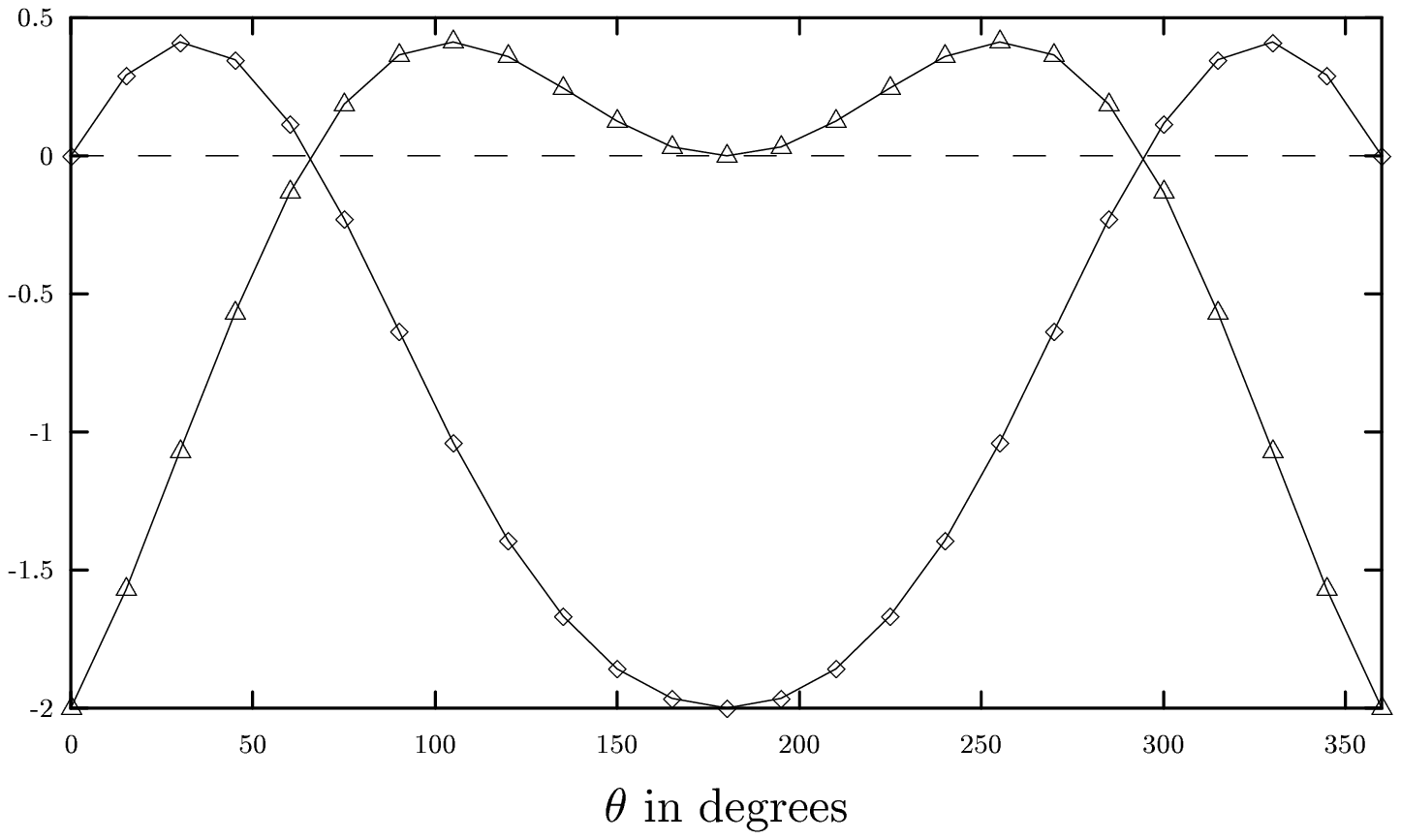}
\vspace*{-5in}
\caption{Variation of squeezing $Q_{x}(\diamond)$ and $Q_{y}(\triangle)$ with respect to
$\theta$ with $\tau=90^{\circ}$ in the upper plot and $\tau=45^{\circ}$ in the lower plot. }
\end{center}
\end{figure}
\begin{figure}[H]
\begin{center}
\vspace*{-1in}
\includegraphics[width=\linewidth]{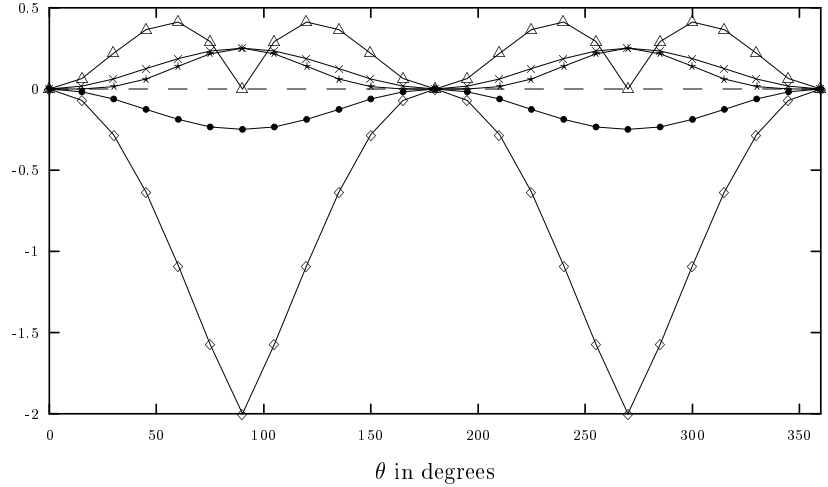}\\
\vspace*{-5in}
\caption{ Variation of spin-spin 
correlations $D^{12}_{x_{1} x_{2}}$  ($\times$), $D^{12}_{y_{1} y_{2}}$  
($\bullet$) and $D^{12}_{z_{1}z_{2}}$ ($\star$).  
 The plot also show squeezing  
in $Q_{x}$ ($\diamond$) 
and $Q_{y}$  ($\triangle$) with respect to 
$\theta$ with $\tau$=90}.
\end{center}
\end{figure}
\section{Summary}
We have looked into the squeezing aspect of a pure bipartite 
state consisting of two spinors. A suitable criterion 
for squeezing of such states 
has been obtained which is a generalization of the squeezing 
criterion for states of sharp spin. While squeezing is established 
in the case of sharp spins due to self correlations, that for a bipartite state 	
occurs due to the presence of both self and mutual correlations.
The existence of mutual 
correlations also implies entanglement. This raises the 
question whether
every entangled state is squeezed. We have shown that all entangled
states of two spinors are squeezed except the singlet state which
is an exception. This state lacks squeezing since it has both
$\langle \vec S_1\rangle=0, \langle \vec S_2\rangle=0$. A direct product
of two spinors, on the other hand has neither self nor mutual 
correlations and hence is never squeezed. However,
if $s_{1}$ or $s_{2}>\frac{1}{2}$, then a direct product state
can indeed possess self correlations and such a bipartite state may 
show squeezing. An example for this could be the direct product 
state of spin 1 squeezed state with a spin $\frac{1}{2}$ state.
These situations indicate that while entanglement stems out from only 
mutual correlations, squeezing arises because due to both of them and when there
is net mean spin value for either of the subsystems.

Our study in this paper gives some justification to some of the claims made
by Kitagawa and Ueda earlier regarding what exactly causes squeezing 
and how squeezed states can be generated. We have shown that  
spin- spin interactions can lead to entangled as well as squeezed states.
    
Further studies on these aspects are under progress where we are also planning to 
analyze the squeezing of mixed states of bipartite systems in which 
there are not only quantum correlations among and within the
spins but also correlations arising due to the nature of statistical 
distribution of the spin assemblies.
\section *{Acknowledgement}
KSM thanks the University of Mysore for financial 
assistance under the minor research project (UGC unassigned grant) entitled 
`Interaction of multi- level quantum systems with external fields'.


\section*{References}    
\begin{thebibliography}{99}
\bibitem{meh1}Mallesh K S, Swarnamala Sirsi, Mahmoud A A Sbaih, Deepak P N and 
Ramachandran G 2000 {\it J. Phys. A: Math. Gen.} {\bf 33} 779
\bibitem{meh2}Mallesh K S, Swarnamala Sirsi, Mahmoud A A Sbaih, Deepak P N and 
Ramachandran G 2000 {\it J. Phys. A: Math. Gen.} {\bf 34} 3293
\bibitem{har}Kimble H J and Walls D F (ed) 1987 {\it J. Opt. Soc.} {\bf B 4} 1450
\nonum Loudon R and Knight P L 1987 {\it J. Mod. Opt.} {\bf 34}
\bibitem{rad}Stoler D 1970 {\it Phys. Rev.} D {\bf 1} 3217
\nonum Yuen H P 1976 {\it Phys. Rev.} A {\bf 13} 2226
\nonum Walls D F 1983 {\it Nature} {\bf 306} 141
\nonum Hollenhorst J N 1979 {\it Phys. Rev.} D {\bf 19} 1669
\nonum Caves C M and Schumaker B L 1985 {\it Phys. Rev.} A {\bf 31} 3068
\nonum Maeda M W, Kumar P and Shapiro K J H 1987 {\it Opt. Lett.}
{\bf 12} 161
\bibitem{ku}Kitagawa M and Ueda M 1993 {\it Phys. Rev.} A {\bf 47} 5138 
\bibitem{puri}Puri R R 1997 {\it Pramana} {\bf 48} 787
\bibitem{wine}Wineland D J, Bollinger J , Itano W M, Moore F L and
Henzen D J 1992 {\it Phys. Rev.} A {\bf 46} 6797 
\bibitem{will} Willil-Hans Steeb and Yorick Hardy {\it Int. J. Mod. Phys. C} {\bf 11} 69
\bibitem{sch}Schwinger J 1965 {\it Quantum Theory in Angular Momentum}
ed L C Biedenharn and H van Dam (New York: Academic) p 230
\bibitem{ram}Ramachandran  G and Mallesh K S  1984 {\it  Nucl. Phys} A {\bf 422} 327 
\end{thebibliography}
\end{document}